# Electrodynamics as a particular case of the most general relativistic force field

STEFANO RE FIORENTIN

*Fiat Group Automobiles, Innovation Department, 200 Corso Agnelli, 10135 Turin, Italy*

**Abstract**. A new approach to classical electrodynamics is presented, showing that it can be regarded as a particular case of the most general relativistic force field. In particular, at first it is shown that the structure of the Lorentz force comes directly from the structure of the three-force transformation law, and that $E$ and $B$ fields can be defined, which in general will depend not only on the space-time coordinate, but also on the velocity of the body acted upon. Then it is proved that these fields become independent from the body velocity if the force field propagates throughout space at the relativistic speed limit $c$. Finally, field sources are introduced by defining them as perturbations of the field itself, obtaining a generalization of Maxwell equations which allow to express the field in terms sources, if these last are known a priori. Electrodynamics follows simply assuming, in addition to the postulate of field propagation at the speed $c$, that scalar sources of the $B$ field are missing and that electric charge is an invariant characteristic of matter, acting both as source of the field and sensing property. The presented approach may have valuable didactic effectiveness in showing that space-time structure, as defined by Special Relativity postulates, is the ultimate responsible of the structure of electrodynamics laws.

## 1. Introduction

In almost every treatise of Electrodynamics and Relativity, Maxwell equations are proved to be covariant under Lorentz transformations, and they are simply postulated as fundamental laws of Nature. Exceptions to this approach are found in Frisch and Wilets [1], and in Tessman [2] who derived Maxwell's equations from Coulomb law; in Schwartz [3] and in Ohanian [4], who showed that the electric field cannot provide a complete description, but that there must be in addition some other field quantities at each point in space which exert a velocity-dependent force. Subsequently Kobe [5] derived Maxwell equations from electrostatics, while Neuenschwander et Turner [6] obtained the same equations by generalizing the laws of magnetostatics. More recently Ohanian [7] derived Maxwell equations from the mere knowledge of the existence of a long-range interaction and assuming, in analogy with Lagrangian mechanics, that the electromagnetic field could be described by a vector field.

We want to present here a new approach to electrodynamics, showing that it can be regarded as a particular case of the most general relativistic force field, if a few particular assumptions are accepted, regarding from the one hand the field propagation speed, and from the other hand the nature and properties of field sources.

## 2. The most general relativistic force field

### 2.1. The four-force and its transformation law

Let us consider, in the context of Special Relativity, an inertial reference system $S'$ moving with a normalized velocity $v$ (throughout the whole paper we shall adopt the convention of normalizing velocities with respect to the speed of light $c$) with respect to another inertial reference system $S$. A general Lorentz transformation without rotation from $S'$ to $S$ implies the following transformation law for a generic four-vector $A^\mu$:

$$A^\mu = \frac{\partial x^\mu}{\partial x'^\alpha} A'^\alpha \qquad (\mu, \alpha = 0,1,2,3) \tag{1}$$

where[6]:

$$\frac{\partial x^\mu}{\partial x'^\alpha} = \begin{vmatrix} \gamma_v & \gamma_v v_x & \gamma_v v_y & \gamma_v v_z \\ \gamma_v v_x & 1+\alpha_v v_x^2 & \alpha_v v_x v_y & \alpha_v v_x v_z \\ \gamma_v v_y & \alpha_v v_y v_x & 1+\alpha_v v_y^2 & \alpha_v v_y v_z \\ \gamma_v v_z & \alpha_v v_z v_x & \alpha_v v_z v_y & 1+\alpha_v v_z^2 \end{vmatrix} \tag{2}$$

In the previous expression we have put:

$$\gamma_v = (1-v^2)^{-1/2} \tag{3}$$

$$\alpha_v = (\gamma_v - 1)/v^2 = \gamma_v^2/(\gamma_v + 1) \tag{4}$$

If we assume:

$$A^\mu = \{a^0, a^1, a^2, a^3\} = \{a^0, \boldsymbol{a}\} \tag{5}$$

in view of expression (2) the equation (1) gives:

$$\boldsymbol{a} = \boldsymbol{a}' + \alpha_v \boldsymbol{v}(\boldsymbol{a}'\cdot\boldsymbol{v}) + \gamma_v a'^0 \boldsymbol{v} \tag{6a}$$

$$a^0 = \gamma_v(a'^0 + \boldsymbol{a}'\cdot\boldsymbol{v}) \tag{6b}$$

In the following we will make use of the normalized four velocity $U^\mu$ of a generic particle, defined by:

$$U^\mu \equiv \frac{1}{c}\frac{dx^\mu}{d\tau} = \frac{1}{c}\frac{dx^\mu}{dt}\frac{dt}{d\tau} = \gamma_u\{1, \boldsymbol{u}\} \tag{7}$$

where $\tau$ is the proper time of the particle, $\boldsymbol{u}$ is its normalized three-velocity and:

$$\gamma_u = (1-u^2)^{-1/2} \tag{8}$$

Starting from the four-velocity, we define the four-momentum of the same particle:

$$P^\mu \equiv cm_0 U^\mu = \left\{\frac{E_{tot}}{c}, \boldsymbol{p}\right\} \tag{9}$$

where $m_0$ is the proper mass of the particle and $\boldsymbol{p}$ its three-momentum. The four-vector $P^\mu$ allows us to build the invariant $P^\mu P_\mu = m_0^2 c^2$ from which the following relationship can be derived:

$$E_{tot}^2 - p^2 c^2 = m_0^2 c^4 \tag{10}$$

In this paper we are interested in the four-force acting on the particle, defined by:

$$F^\mu \equiv \frac{dP^\mu}{d\tau} = \frac{dP^\mu}{dt}\frac{dt}{d\tau} = \gamma_u \frac{dP^\mu}{dt} \tag{11}$$

For a purely mechanical force (a force that does not alter the proper mass of the bodies acted upon by it), the four-force takes the form:

$$F^\mu = \gamma_u \{(\boldsymbol{f} \cdot \boldsymbol{u}), \boldsymbol{f}\} \tag{12}$$

where $\boldsymbol{f}$ is the three-force, defined by:

$$\boldsymbol{f} \equiv \frac{d\boldsymbol{p}}{dt} = \frac{d\boldsymbol{p}}{dt'} / \frac{dt}{dt'} \tag{13}$$

In view of equations (6) and (9), the two terms on the right in equation (13) are given by:

$$\frac{d\boldsymbol{p}}{dt'} = \frac{d}{dt'}\left[\boldsymbol{p}' + \alpha_v \boldsymbol{v}(\boldsymbol{p}' \cdot \boldsymbol{v}) + \gamma_v \boldsymbol{v} \frac{E'_{tot}}{c}\right] \tag{14}$$

$$\frac{dt}{dt'} = \frac{d}{dt'}\left[\gamma_v \left(t' + \frac{\boldsymbol{x}' \cdot \boldsymbol{v}}{c}\right)\right] \tag{15}$$

Performing the derivatives into (14) and (15), taking into account expression (12), and substituting the result into (13), we get the transformation law of the three-force [8]:

$$\boldsymbol{f} = \frac{\boldsymbol{f}' + \alpha_v \boldsymbol{v}(\boldsymbol{f}' \cdot \boldsymbol{v}) + \gamma_v \boldsymbol{v}(\boldsymbol{f}' \cdot \boldsymbol{u}')}{\gamma_v (1 + \boldsymbol{u}' \cdot \boldsymbol{v})} \tag{16}$$

It is particularly interesting that the transformation law of the force from a system of inertia $S'$ to another system of inertia $S$ has an expression explicitly depending not only on the velocity $\boldsymbol{v}$ of $S'$ with respect to $S$, but also on the velocity $\boldsymbol{u}'$ of the particle acted upon by the force in the reference frame $S'$. This dependence of the force on particle velocity is a direct consequence of space-time structure and, therefore, it can be ascribed to the fundamental postulates of Special Relativity. The force $\boldsymbol{f}'$, in its turn, may depend on particle velocity by virtue of the same transformation effect with respect to a third reference frame $S''$, and so on. One could think that there should possibly be a reference frame in which the force does not depend on particle velocity. But this could not happen, due to the very fundamental properties of the force, which reside in the way it is transmitted through space. In general, we may therefore say that, in a generic reference frame, the force $\boldsymbol{f}$ shows a dependence on particle velocity due to two causes: the intrinsic nature of the force itself and the structure of its transformation law.

*2.2. Decomposition of the force dependence on body velocity*

In this paper we want to study in detail this twofold dependence on body velocity of the force, with the aim of separating the two effects by putting in an explicit form the component due to the structure of the transformation law (16). We recognize that the structure of this explicit dependence of force on body velocity has to be invariant: in other terms, if formulated to represent the dependence of $f'$ on $u'$, when substituted into (16) must give to $f$ a similar explicit dependence on $u$. First of all, we observe that since in principle $f'$ could not depend on $u'$ (this may occur in very particular situations, depending both on the nature of the force and on the choice of the reference system $S'$), equation (16) itself must show the functional dependence of $f$ on $u$ we are looking for, provided we substitute for $u'$ its expression in terms of $u$. Let us therefore accomplish this substitution.

The reverse of equations (6a) and (6b), in the case in which the four-vector (5) corresponds to the four-velocity (7), become:

$$\gamma_{u'} u' = \gamma_u [u + \alpha_v v (u \cdot v) - \gamma_v v] \tag{17a}$$

$$\gamma_{u'} = \gamma_u \gamma_v (1 - u \cdot v) \tag{17b}$$

Dividing (17a) by (17b) we get [8]:

$$u' = \frac{u + \alpha_v v (u \cdot v) - \gamma_v v}{\gamma_v (1 - u \cdot v)} \tag{18}$$

Now we can proceed with the substitution of $u'$ in terms of $u$ into equation (16). The denominator $\mathcal{D}$ of expression (16), after the substitution of (18), becomes:

$$\mathcal{D} = \gamma_v (1 + u' \cdot v) = \frac{1}{\gamma_v (1 - u \cdot v)} \tag{19}$$

The same substitution in the numerator $\mathcal{N}$ of (16) gives:

$$\mathcal{N} = f' + \alpha_v v (f' \cdot v) + \gamma_v v (f' \cdot u') = \frac{\gamma_v [f' + v(f' \cdot u) - f'(u \cdot v)] - \alpha_v v(f' \cdot v)}{\gamma_v (1 - u \cdot v)} \tag{20}$$

Dividing equation (20) by equation (19) we get the new expression of the force transformation law, where the velocity $u'$, measured with respect to the system $S'$, has been replaced by the velocity $u$ measured with respect to $S$:

$$f = \gamma_v [f' + v(u \cdot f') - f'(u \cdot v)] - \alpha_v v(f' \cdot v) \tag{21}$$

Making use of the vector identity:

$$v(u \cdot f') - f'(u \cdot v) = u \times (v \times f') \tag{22}$$

equation (21) can be rewritten in the form:

$$f = [\gamma_v f' - \alpha_v v(f' \cdot v)] + u \times [\gamma_v (v \times f')] \tag{23}$$

We now assume to be in one of the aforementioned particular cases in which $f'$ does not depend on $u'$, even if in general it will depend on particle space-time coordinates of the $S'$ frame: $f' = f'(t', x')$. This dependence allows us to speak of $f'$ as of a *force field*. Since $t', x'$ can be expressed in terms of $t, x$, we are allowed to put:

$$\overline{E}(t, x) = \gamma_v f' - \alpha_v v (f' \cdot v) \tag{24a}$$

$$c\overline{B}(t, x) = \gamma_v (v \times f') \tag{24b}$$

where $\overline{E}(t, x)$ and $c\overline{B}(t, x)$ can in their turn be regarded as fields. (We use the quantity $c\overline{B}$ instead of $\overline{B}$ so that, if reference is made to the non normalized $u$ velocity, the constant $c$ simplifies out).

In terms of definitions (24a) and (24b), equation (23) can be rewritten in the form:

$$f(t, x, u) = \overline{E}(t, x) + u \times c\overline{B}(t, x) \tag{25}$$

This equation represents the sought-after prototypal dependence of $f$ on velocity $u$ due to transformation effects only. If we remove the hypothesis that $f'$ does not depend on $u'$, in view of the assumed general validity of the structure of expression (25), we have simply to introduce a dependence on velocity $u$ into fields $\overline{E}(t, x)$ and $c\overline{B}(t, x)$:

$$f(t, x, u) = \overline{E}(t, x, u) + u \times c\overline{B}(t, x, u) \tag{26a}$$

Now, according to our hypothesis, this expression has to be invariant in form, and therefore in the $S'$ frame of reference must take the form:

$$f'(t', x', u') = \overline{E}'(t', x', u') + u' \times c\overline{B}'(t', x', u') \tag{26b}$$

If relations (26a) and (26b) are consistent, the dependence of the force $f$ on particle velocity due to transformation effects only, must be described by the cross product $u \times$, whereas the terms $\overline{E}(t, x, u)$ and $c\overline{B}(t, x, u)$ describe exclusively the dependence on $u$ due to the intrinsic nature of the force. To prove this, we have to substitute (26b) into (16), and check if the result takes the form of (26a). We get:

$$f = \frac{\overline{E}' + \alpha_v v(\overline{E}' \cdot v) + \gamma_v v(\overline{E}' \cdot u')}{\gamma_v(1 + u' \cdot v)} + \frac{(u' \times c\overline{B}') + v[\alpha_v (u' \times c\overline{B}') \cdot v]}{\gamma_v(1 + u' \cdot v)} \tag{27}$$

The first term in the right member of (27) (let us denote it by $f^{(1)}$) is similar to equation (16), where simply $f'$ has been replaced by $\overline{E}'$. The substitution into it of $u'$ in terms of $u$ follows the same path that led us from equation (16) to equation (23), giving:

$$f^{(1)} = [\gamma_v \overline{E}' - \alpha_v v(\overline{E}' \cdot v)] + u \times [\gamma_v (v \times \overline{E}')] \tag{28}$$

Let us now take into account the second term in the right member of equation (27) (denote it by $f^{(2)}$), and proceed with the same substitution. We get:

$$f^{(2)} = (u \times c\overline{B}') + (v \times c\overline{B}')(\alpha_v u \cdot v - \gamma_v) + v[\alpha_v u \cdot (c\overline{B}' \times v)] \tag{29}$$

The vector identity:

$$(v \times c\overline{B}')(u \cdot v) = u \times [v^2 c\overline{B}' - v(v \cdot c\overline{B}')] - v[u \cdot (c\overline{B}' \times v)] \tag{30}$$

allows us to rewrite equation (29) in the form:

$$f^{(2)} = \gamma_v(u \times c\overline{B}') - \gamma_v(v \times c\overline{B}') - \alpha_v(u \times v)(v \cdot c\overline{B}') \tag{31}$$

Adding equations (28) and (31) we get the following final expression for the force $f$:

$$f = [\gamma_v \overline{E}' - \alpha_v v(v \cdot \overline{E}') - \gamma_v(v \times c\overline{B}')] + u \times [\gamma_v c\overline{B}' - \alpha_v v(v \cdot c\overline{B}') + \gamma_v(v \times \overline{E}')] \tag{32}$$

This equation assures us about the general validity of expressions (26), and, at the same time, gives us the transformation laws for $\overline{E}$ and $c\overline{B}$:

$$\overline{E} = \gamma_v \overline{E}' - \alpha_v v(v \cdot \overline{E}') - \gamma_v(v \times c\overline{B}') \tag{33a}$$

$$c\overline{B} = \gamma_v c\overline{B}' - \alpha_v v(v \cdot c\overline{B}') + \gamma_v(v \times \overline{E}') \tag{33b}$$

We recognize that previous equations (24a) and (24b) were simply the particular case of these last equations, in which $\boldsymbol{f'}$ does not depend on $\boldsymbol{u'}$, and therefore, on the basis of (26b), $c\overline{\boldsymbol{B}}' = \boldsymbol{0}$ and $\boldsymbol{f'}(t,\boldsymbol{x}) = \overline{\boldsymbol{E}}'(t,\boldsymbol{x})$,

We have thus reached our goal, having confined into the components $\overline{\boldsymbol{E}}(t,\boldsymbol{x},\boldsymbol{u})$ and $c\overline{\boldsymbol{B}}(t,\boldsymbol{x},\boldsymbol{u})$ the possible dependence on velocity $\boldsymbol{u}$ of the force due to its intrinsic nature (that is, essentially, to the way in which it is transmitted through space), while expressions (26) account for the additional dependence of the force on particle velocity due to transformation effects only.

We give to expressions (26) the name of *generalized Lorentz force* after Hendrik Antoon Lorentz, following a tradition well established in electrodynamics. It is remarkable the fact that these expressions are of general validity, and apply to any kind of force, even if we are used to see them applied to electromagnetic forces.

*2.3. En example of force field*

To better figure out how things go, just think, for example, of a force which is transmitted by an isotropic flux of tiny particles (vector particles) of rest mass $m_p$ originating from a point source at rest in the reference frame $S$, propagating in the empty space with radial velocity $\boldsymbol{w}$, and impinging on our probe, which is moving at velocity $\boldsymbol{u}$ in $S$ and which shows a certain capture cross section for the vector particles of the flux (we assume this absorption cross section to be an invariant). It is clear that the force on the probe will depend, among the other things, on its velocity $\boldsymbol{u}$, even in the reference frame $S$ where the source is at rest. Let us determine this dependence. The force experienced by the probe is given by the momentum variation of the vector particles impinging on it in the unit of time. The momentum $\boldsymbol{p}_p$ of each vector particle emerging from the source is:

$$\boldsymbol{p}_p = \left(1 - w^2\right)^{-1/2} m_p c\, \boldsymbol{w} = \gamma_w m_p c\, \boldsymbol{w} \tag{34}$$

or:

$$\boldsymbol{p}_p = E_p \boldsymbol{w}/c \tag{35}$$

where $E_p = \gamma_w m_p c^2$ is the total energy of the vector particle. The final momentum of the same vector particle after the collision with the probe (we assume the mass $m_p$ negligible with respect to the mass of the probe) is:

$$\boldsymbol{p}_f = \gamma_u m_p c\, \boldsymbol{u} = \frac{E_p \gamma_u}{c \gamma_w} \boldsymbol{u} \tag{36}$$

If $I$ is the total number of vector particles emitted by the source per unit of time, and $\Sigma_a$ is the invariant absorption cross section of the probe, the number $N$ of vector particles impinging on it in the unit of time is:

$$N = \frac{I \Sigma_a}{4\pi r^2} \tag{37}$$

where $r$ is the distance of the probe from the source. If the vector particles decay during their travel, the number of those which succeed in impinging on the probe is just the number $N$ reduced by a factor $e^{-\lambda r}$. Using equations (35), (36) and (37), we can therefore express the force acting on the probe in the form:

$$f = \frac{I \Sigma_a e^{-\lambda r}}{4\pi r^2} \frac{E_p}{c} \left( w - \frac{\gamma_u}{\gamma_w} u \right) \tag{38}$$

We can observe that if $w = u$ the force vanishes, as expected. If we put the point source in the origin of the coordinate system of $S$, we have $r = |x|$ and $w = wx/r$, so that:

$$f(x,u) = \frac{I \Sigma_a e^{-\lambda r}}{4\pi r^2} \frac{E_p}{c} \left( \frac{w}{r} x - \frac{\gamma_u}{\gamma_w} u \right) \tag{39}$$

Comparing this expression with equation (26a), we infer that $\overline{E}(x,u) \equiv f(x,u)$ and $c\overline{B} \equiv 0$.

We can now get the expression of the force in a generic frame of reference (call it $S'$), moving with velocity $v$ with respect to $S$, by means of equation (26b) with $\overline{E}'$ and $c\overline{B}'$ given by the inverse of equations (33). (These can be obtained by simply interchanging $\overline{E}$ and $c\overline{B}$ with $\overline{E}'$ and $c\overline{B}'$ respectively, and changing the sign of $v$). Of course, once substituted the expression of $\overline{E}(x,u)$, given by equation (39), into the inverse of equations (33), we have to express $u$ in terms of $u'$ and $x$ in terms of $(x', t')$ to get consistent relations for $\overline{E}'$ and $c\overline{B}'$.

### 3. Force fields that propagate at the speed of light

*3.1 Postulates*

We now want to concentrate on the particular cases in which the fields $\overline{E}$ and $c\overline{B}$ do not depend on velocity $u$ of the sensing body, but only on its space-time coordinates. Referring to the previous example, we can easily prove that if the vector "particles" emitted by the source and responsible of the force field, move at the speed $c$, the dependence on velocity $u$ disappears in equation (39). To this end, we have to take into account that when $w = 1$, from equation (35) we get $E_p = p_p c$ while from equation (10) we deduce that these "particles" have no rest mass. Furthermore they cannot decay while travelling into space since their proper time does not roll by: therefore $\lambda = 0$ in equation (39). Taking the limit of this equation for $w \to 1$, we therefore obtain:

$$f(x) = \frac{I \Sigma_a E_p}{4\pi r^3 c} x \tag{40}$$

This equation can be regarded as a model of how things go when the force is transmitted by "particles" that originate from the source and travel through the empty space at the speed $c$. In the following we want to confine our analysis to these particular cases of force fields, looking for the general expressions that govern the interaction. In doing this we assume that the "vector particles", whatever they are, originate only from material sources. This assumption essentially rules out the case of non linear fields, where the field itself plays the role of a source. This happens for gravitation, whose source is energy, and where the field itself, having an energy, plays the role of a source.

We therefore assume the following two postulates:

**Postulate n.1:**
**The force is transmitted trough the empty space at the speed *c* and originates exclusively from material sources.**

This postulate assures the linearity of the field and rules out any dependence on particle velocity of the fields $\overline{E}$ and $c\overline{B}$.

**Postulate n.2:**
**The property of a body that makes it sensitive to the force field is described by an invariant scalar, independent from the velocity of the body.**

We will call *charge* this invariant property of matter and will indicate it by *q*. This name comes after the *electric charge*, which - as we will see - is the most relevant example of property of this kind. In the previous example this property was described by the invariant absorption cross section $\Sigma_a$.

At this point, it is useful to normalize the force field experienced by a body with respect to its charge *q*, getting a representation of the force field as perceived by a unitary charge. To this end we define:

$$E \equiv \frac{\overline{E}}{q}; \quad B \equiv \frac{\overline{B}}{q} \tag{41}$$

The vector fields *E* and *cB* give us a complete representation of the force field, independently from the bodies that, placed into it, experience the force. For this reason we call *E* and *cB* *force field vectors*. By virtue of equations (41), we can re-write the transformation equations (33) in the form:

$$E = \gamma_v E' - \alpha_v v(v \cdot E') - \gamma_v (v \times cB') \tag{42a}$$

$$cB = \gamma_v cB' - \alpha_v v(v \cdot cB') + \gamma_v (v \times E') \tag{42b}$$

Knowing that, thanks to Postulate 1, *E* and *cB* are only functions of space-time coordinates, we observe that these transformation laws indissolubly link fields *E* and *cB*, exactly as Lorentz transformations indissolubly link space and time. A field exclusively described by *E* (or by *cB*) may only exist in a particular reference frame: it is sufficient to change frame of reference to get a field *cB* associated to field *E* (or vice versa). Therefore, what only has a physical reality (independently from the frame of reference) is the ensemble of *E* and *cB* fields, not *E* or *cB* separately, as in kinematics what has a physical reality is space-time, not space and time separately.

In view of definitions (41) and Postulate n.1, the expression (26a) of the Lorentz force becomes:

$$f(t,x,u) = q[E(t,x) + u \times cB(t,x)] \tag{43}$$

*3.2. Covariant formulation of the Lorentz force*

The next step of our analysis is to give covariant form to equations (26), in view of their proved general validity. First of all, in place of the three-force we must take into account the four-force (12). Then, if we define the quantities $F^{\mu\nu}$:

$$F^{\mu\nu} \equiv \begin{bmatrix} 0 & -E_x & -E_y & -E_z \\ E_x & 0 & -cB_z & cB_y \\ E_y & cB_z & 0 & -cB_x \\ E_z & -cB_y & cB_x & 0 \end{bmatrix} \tag{44}$$

and take the product of $F^{\mu\nu}$ with the covariant four-velocity $U_\nu$, we can easily verify that the expression:

$$F^{\mu} = qF^{\mu\nu}U_{\nu} \tag{45}$$

corresponds to equation (43). Since $F^{\mu}$ and $U_{\nu}$ are tensors, and in particular $U_{\nu}$ is an arbitrary tensor, the quotient law assures us that the quantities $F^{\mu\nu}$ transform like an anti-symmetric tensor. This one enjoys the valuable property of containing the whole description of the force field, having as its components the vector fields $\boldsymbol{E}(t,\boldsymbol{x})$ and $c\boldsymbol{B}(t,\boldsymbol{x})$; it is therefore referred to as the *force field tensor*. Its transformation law:

$$F^{\mu\nu} = \frac{\partial x^{\mu}}{\partial x'^{\alpha}} \frac{\partial x^{\nu}}{\partial x'^{\beta}} F'^{\alpha\beta} \tag{46}$$

gives us exactly the transformations laws (42), as it may be easily verified using expression (2).

*3.3. The scalar field sources*

If a force field exists in space-time it is because there is something that generates it: that is, there are *sources*. These can only be perceived through the field they generate: a more direct perception is not possible. It follows that sources have to be *defined* in terms of perturbations of the field. We now want to come to a proper definition of the sources of a linear force field that propagates at the speed *c*. First of all, we recognize that a proper definition of sources must assure them to be invariant: the description of the sources contained in every given volume must be independent of the reference frame. Then, we know from vector field theory that divergence and curl are two operators that quantify perturbations of a field and are therefore suitable to build the definitions of sources of a force field.

Let us start considering divergence. Since we are looking for an invariant, we must take into account the divergence of the four-force. We therefore define:

$$\omega^{(S)}(\boldsymbol{u}) \equiv \varepsilon_0 \partial_{\mu} F^{\mu}(\boldsymbol{u}) \tag{47}$$

where $^{(S)}$ stands for "scalar", and $\varepsilon_0$ is an appropriate constant, introduced to account for specific units of measure for $\omega^{(S)}$. We call $\omega^{(S)}$ *scalar source density of the force field*. We observe that, even if $\omega^{(S)}$ depends on $\boldsymbol{u}$, it is nonetheless invariant.

At this point, in order to put in an explicit form the dependence on $\boldsymbol{u}$ of $\omega^{(S)}(\boldsymbol{u})$, we resort to the field tensor $F^{\mu\nu}$. Using the equation (45) and taking into account that velocity $\boldsymbol{u}$ is chosen at will and therefore can be considered a constant, we get:

$$\partial_{\mu} F^{\mu}(\boldsymbol{u}) = q \partial_{\mu}[F^{\mu\nu} U_{\nu}(\boldsymbol{u})] = q[\partial_{\mu} F^{\mu\nu}] U_{\nu}(\boldsymbol{u}) \tag{48}$$

Substituting this expression into the definition (47) we get:

$$\omega^{(S)}(\boldsymbol{u}) \equiv q[\varepsilon_0 \partial_{\mu} F^{\mu\nu}] U_{\nu}(\boldsymbol{u}) \tag{49}$$

This expression makes explicit the dependence on $\boldsymbol{u}$ of the scalar source density, confining it into the term $U_{\nu}(\boldsymbol{u})$, since the field tensor $F^{\mu\nu}$ does not depend on $\boldsymbol{u}$. We now define:

$$J^{(S)\nu} \equiv \varepsilon_0 \partial_{\mu} F^{\mu\nu} \tag{50}$$

The four-vector $J^{(S)^V}$ can be regarded as the scalar source of the field tensor, and it is independent from the state of motion of any body subjected to the field. Substituting (50) into (49) we get:

$$\omega^{(S)}(\boldsymbol{u}) = qJ^{(S)^V} U_V(\boldsymbol{u}) \tag{51}$$

The four-vector $J^{(S)^V}$ can be thought of as made up of the components:

$$J^{(S)^V} = \left\{ \rho^{(S)}, \frac{\boldsymbol{j}^{(S)}}{c} \right\} \tag{52}$$

so that equation (51) can be put in the form:

$$\omega^{(S)}(\boldsymbol{u}) = q\gamma_u \left[ \rho^{(S)} - \boldsymbol{u} \cdot \frac{\boldsymbol{j}^{(S)}}{c} \right] \tag{53}$$

This is the final result of our process of definition of the scalar sources of the force field, where, being guided by the decomposition (45), we got a corresponding decomposition – represented by equation (51) - of the scalar source density, where the dependence on velocity $\boldsymbol{u}$ has been made explicit. This allowed us to achieve - with equation (50) - a definition of the sources directly responsible for the force field tensor.

Having completed the task of getting a proper definition of what we called scalar field sources under the assumption that the field is given, we now reverse the perspective assuming that scalar sources, as previously defined, are given, and face the problem of finding the force field.

On the basis of equation (50) we have:

$$\partial_\mu F^{\mu\nu} = \frac{1}{\varepsilon_0} J^{(S)^V} \tag{54}$$

where, using definition (44):

$$\partial_\mu F^{\mu\nu} = \left\{ \nabla \cdot \boldsymbol{E}, (\nabla \times c\boldsymbol{B}) - \frac{1}{c}\frac{\partial \boldsymbol{E}}{\partial t} \right\} \tag{55}$$

Substituting this expression into (54) and making use of the representation (52), we get:

$$\nabla \cdot \boldsymbol{E} = \frac{1}{\varepsilon_0} \rho^{(S)} \tag{56a}$$

$$\nabla \times c\boldsymbol{B} - \frac{1}{c}\frac{\partial \boldsymbol{E}}{\partial t} = \frac{1}{\varepsilon_0} \frac{\boldsymbol{j}^{(S)}}{c} \tag{56b}$$

These equations are not sufficient to allow the determination of the six unknown components of the field, since they only contain three independent equations. In fact, equations (56) or (54) are not independent by virtue of the following relation:

$$\partial_\nu \partial_\mu F^{\mu\nu} = 0 \tag{57}$$

which comes from the anti-symmetry of the field tensor. Equation (57), in view of (54), implies:

$$\partial_\nu J^{(S)^V} = 0 \tag{58a}$$

or:

$$\nabla \cdot \boldsymbol{j}^{(S)} + \frac{\partial \rho^{(S)}}{\partial t} = 0 \tag{58b}$$

These relationships express the continuity equation of scalar sources. It is therefore natural to call $\rho^{(S)}$ *scalar source density*, and the three-vector $\boldsymbol{j}^{(S)}$ *current density of scalar sources*, thanks to the meaning they assume in the previous conservation equation. Thanks to these definitions, we call $J^{(S)\nu}$ *four-current density of scalar sources*.

We conclude that, since in the real world the field is necessarily determined by its sources, and since the same must happen in our mathematical model, there must be other field equations, maybe linking the field to other sources, described by the curl of the field tensor. Let us investigate further.

*3.4. The dual field tensor and vortex sources*

Let us now take into account the tensor $F^*_{\kappa\lambda}$ dual to $F^{\mu\nu}$ defined by:

$$F^*_{\kappa\lambda} \equiv \frac{1}{2} \varepsilon_{\kappa\lambda\mu\nu} F^{\mu\nu} \tag{59}$$

where $\varepsilon_{\kappa\lambda\mu\nu}$ is the permutation tensor, anti-symmetric in all four indices. We therefore have, by definition:

$$F^*_{\kappa\lambda} \equiv \begin{bmatrix} 0 & -cB_x & -cB_y & -cB_z \\ cB_x & 0 & -E_z & E_y \\ cB_y & E_z & 0 & -E_x \\ cB_z & -E_y & E_x & 0 \end{bmatrix} \tag{60}$$

Multiplying equation (59) by $\varepsilon^{\alpha\beta\kappa\lambda}/2$ we recognize that $F^{\mu\nu}$ is in its turn dual to $F^*_{\kappa\lambda}$:

$$F^{\mu\nu} = \frac{1}{2} \varepsilon^{\mu\nu\kappa\lambda} F^*_{\kappa\lambda} \tag{61}$$

Substituting expression (61) into equation (50) we get:

$$J^{(S)\nu} = \frac{\varepsilon_0}{2} \varepsilon^{\mu\nu\kappa\lambda} \partial_\mu F^*_{\kappa\lambda} \tag{62}$$

This equation states that $J^{(S)\nu}$, apart from the constant $\varepsilon_0$, is the curl of the dual field tensor $F^*_{\kappa\lambda}$, and therefore it can also be regarded as the *vortex source* of this same tensor. At this point we can argue that even the dual field tensor may have its own scalar sources, that will play the role of vortex sources with respect to the field tensor $F^{\mu\nu}$. In analogy to (50) we therefore define:

$$J^{(V)\lambda} \equiv \varepsilon_0 \partial_\kappa F^{*\kappa\lambda} \tag{63}$$

where $^{(V)}$ stands for "vortex". Substituting the contravariant form of equation (59) into equation (63) we can verify that the curl of the field tensor is actually given by $J^{(V)\lambda}$:

$$\frac{1}{2} \varepsilon^{\kappa\lambda\mu\nu} \partial_\kappa F_{\mu\nu} = \frac{1}{\varepsilon_0} J^{(V)\lambda} \tag{64}$$

This is the equation we were looking for, in order to make the determination of the force field tensor possible from the knowledge of the sources. In fact it just contains the three missing equations. We have in fact to take into account that, as for the field tensor $F^{\mu\nu}$, the following relation holds for the dual field tensor:

$$\partial_\lambda \partial_\kappa F^{*\kappa\lambda} = 0 \tag{65}$$

which implies:

$$\partial_\lambda J^{(V)\lambda} = 0 \tag{66}$$

We conclude that each of the two fields $F^{\mu\nu}$ and $F^{*\kappa\lambda}$ has two kinds of sources: from one side the *scalar sources*, linked to the divergence of the tensor, and from the other side the *vortex sources*, linked to the curl of the tensor. The scalar sources of $F^{\mu\nu}$ are the vortex sources of $F^{*\kappa\lambda}$ and vice versa. The duality between $F^{\mu\nu}$ and $F^{*\kappa\lambda}$ is reflected in the duality between the vector fields $E$ and $cB$. We can express equation (64) in terms of these vector fields. To this end, let us put:

$$J^{(V)} = \left\{ \rho^{(V)}, \frac{\boldsymbol{j}^{(V)}}{c} \right\} \tag{67}$$

where $\rho^{(V)}$ will be called *vortex source density*, and $\boldsymbol{j}^{(V)}$ *current density of vortex sources*, so that:

$$\nabla \cdot c\boldsymbol{B} = -\frac{1}{\varepsilon_0} \rho^{(V)} \tag{68a}$$

$$\nabla \times \boldsymbol{E} + \frac{1}{c}\frac{\partial c\boldsymbol{B}}{\partial t} = \frac{1}{\varepsilon_0}\frac{\boldsymbol{j}^{(V)}}{c} \tag{68b}$$

Once the field sources $J^{(S)\nu}$ and $J^{(V)\lambda}$ are specified, equations (54) and (64) allow the determination of the field tensor $F^{\mu\nu}$ or, equivalently, equations (56) and (68) allow the determination of the vector fields $\boldsymbol{E}$ and $c\boldsymbol{B}$. Subsequently, by means of equations (43) or (45), we can get the force acting on a particle of charge $q$ in any space and time position.

It is useful to note that even if either $J^{(S)\nu}$ or $J^{(V)\lambda}$ is everywhere identically zero, the tensor field, and therefore $\boldsymbol{E}$ and $c\boldsymbol{B}$ fields, can be completely determined. In a certain sense, the fact that both $J^{(S)\nu}$ and $J^{(V)\lambda}$ are non vanishing four-vectors can be regarded as a kind of redundancy. We'll see that electrodynamics does not possess this redundancy. But, in general, there is no reason why one of the two sources should be identically zero everywhere: the general linear force field propagating at speed $c$ has both sources.

### 3.5. Discrete sources

Up to now we have taken into account only distributed sources, whose description is given by the four-vectors $J^{(S)\nu}$ and $J^{(V)\lambda}$. On the contrary, referring to the property that makes matter sensitive to the field, we have taken into account only discrete charges, for which we postulated the invariance. Of course, discrete sources are possible, and we want to analyze their transformation properties. Let us start from the scalar sources. From equations (6) we know that they transform according to the laws:

$$\boldsymbol{j}^{(S)} = \boldsymbol{j'}^{(S)} + \boldsymbol{v}\left[\alpha_v(\boldsymbol{j'}^{(S)} \cdot \boldsymbol{v}) + \gamma_v c\rho'^{(S)}\right] \tag{69a}$$

$$\rho^{(S)} = \gamma_v\left[\rho'^{(S)} + \frac{\boldsymbol{j'}^{(S)}}{c} \cdot \boldsymbol{v}\right] \tag{69b}$$

The discrete source $dQ'^{(S)}$ contained into the small volume $dV'$ at rest in the reference frame $S'$ is given by:

$$dQ'^{(S)} = \rho'^{(S)} dV' \tag{70}$$

Let us suppose that this is the only scalar source in the volume $dV'$, so that, from (69b):

$$\rho^{(S)} = \gamma_v \rho'^{(S)} \tag{71}$$

Since $dV = dV'/\gamma_v$, from equation (71) we get:

$$\rho^{(S)} dV = \rho'^{(S)} dV' \quad \text{or} \quad dQ^{(S)} = dQ'^{(S)} \tag{72}$$

It follows that discrete scalar sources are invariant, exactly as charges are. The same reasoning can be applied to vortex sources, coming to the result that also discrete vortex sources are invariant.

In conclusion at this point we have three invariant properties of matter: two related to sources (scalar and vortex ones), and one related to the property that makes matter sensitive to the field generated by sources (property that we called "charge").

*3.6. The force field equations*

We now come back to consider the force field $\boldsymbol{f}$ and try to link it directly to the field sources $J^{(S)\nu}$ and $J^{(V)\lambda}$ without using its decomposition into the $\boldsymbol{E}$ and $c\boldsymbol{B}$ fields. First of all, we consider scalar sources making reference to expression (53). Substituting this one into equation (47), and taking into account (12), we get the following field equation for the force experienced by a body of charge $q$:

$$\nabla \cdot \boldsymbol{f} + \frac{\boldsymbol{u}}{c} \cdot \frac{\partial \boldsymbol{f}}{\partial t} = \frac{q}{\varepsilon_0}\left[\rho^{(S)} - \boldsymbol{u} \cdot \frac{\boldsymbol{j}^{(S)}}{c}\right] \tag{73}$$

This scalar equation does not allow the determination of the three components of the force field. We have to look for another equation, possibly a vector one, containing also the vortex sources.

Taking the curl of Lorentz expression (43) and using equations (68) we are led to the following expression:

$$\nabla \times \boldsymbol{f} + q\frac{D\boldsymbol{B}}{Dt} = \frac{q}{\varepsilon_0}\left[\frac{\boldsymbol{j}^{(V)}}{c} - \boldsymbol{u}\rho^{(V)}\right] \tag{74}$$

where:

$$\frac{D\boldsymbol{B}}{Dt} = (c\boldsymbol{u} \cdot \nabla)\boldsymbol{B} + \frac{\partial \boldsymbol{B}}{\partial t} \tag{75}$$

is the total time derivative (or substantial derivative) of the **B** field as perceived by the particle moving at velocity *cu*. Expression (74) is not yet the result we want, since it contains the **B** field. However we know how to express $(\nabla \times c\boldsymbol{B})$ in terms of the **E** field thanks to (56b), so we are led to take the curl of equation (74). Thanks to the commutation properties of curl and substantial derivative, we have:

$$\nabla \times (\nabla \times \boldsymbol{f}) + q \frac{D}{Dt}(\nabla \times \boldsymbol{B}) = \frac{q}{\varepsilon_0}\left[\nabla \times \frac{\boldsymbol{j}^{(V)}}{c} - \nabla \times \boldsymbol{u}\rho^{(V)}\right] \tag{76}$$

Making use of (56b), of some product rules, and of equation (68b), we get the following expression:

$$\nabla^2 \boldsymbol{f} - \frac{1}{c^2}\frac{\partial^2 \boldsymbol{f}}{\partial t^2} =$$

$$= \frac{q}{\varepsilon_0}\left[\nabla \rho^{(S)} + \frac{1}{c^2}\frac{\partial \boldsymbol{j}^{(S)}}{\partial t} - \nabla \times \frac{\boldsymbol{j}^{(V)}}{c}\right] - \frac{q}{\varepsilon_0}\boldsymbol{u} \times \left[\nabla \rho^{(V)} + \frac{1}{c^2}\frac{\partial \boldsymbol{j}^{(V)}}{\partial t} + \nabla \times \frac{\boldsymbol{j}^{(S)}}{c}\right] \tag{77}$$

This is the vector field equation for the force field *f* we were looking for, even if it is of the second order: that is the price we have paid for having a decoupled equation. Together with equation (73), it allows the determination of the force field on the basis of the knowledge of field sources and of particle velocity *u*, without resorting to **E** and *c***B** fields.

*3.7. Radiation*

We observe that in the regions of space where all sources are zero, that is:

$$J^{(S)\mu} = 0 \text{ and } J^{(V)\mu} = 0 \tag{78}$$

equation (77) becomes:

$$\nabla^2 \boldsymbol{f} - \frac{1}{c^2}\frac{\partial^2 \boldsymbol{f}}{\partial t^2} = 0 \tag{79}$$

This is a wave equation, describing a wave-like propagation of the field *f* at the speed *c*. This is an expected result, in view of Postulate n.1.

We are interested in investigating if the same radiation property applies also to the field components **E** and *c***B** in the regions where no sources are present. Equations (54) and (64) tell us that in these regions the field tensor becomes both solenoidal and irrotational:

$$\partial_\mu F^{\mu\nu} = 0 \tag{80}$$

$$\varepsilon_{\kappa\lambda\mu\nu}\partial^\kappa F^{\mu\nu} = 0 \tag{81a}$$

Equation (81a) can also be written in the form:

$$\partial^\beta F^{\mu\nu} + \partial^\mu F^{\nu\beta} + \partial^\nu F^{\beta\mu} = 0 \tag{81b}$$

Taking the gradient $\partial_\beta$ of this last equation, we get:

$$\partial_\beta\partial^\beta F^{\mu\nu} + \partial_\beta\partial^\mu F^{\nu\beta} + \partial_\beta\partial^\nu F^{\beta\mu} = 0 \tag{82}$$

Now, the second and the third terms are zero by virtue of (80) and of the reversibility of derivative order, so that equation (82) reduces to:

$$\partial_\beta \partial^\beta F^{\mu\nu} = 0 \tag{83a}$$

or:

$$\begin{cases} \nabla^2 \boldsymbol{E} - \dfrac{1}{c^2} \dfrac{\partial^2 \boldsymbol{E}}{\partial t^2} = 0 \\ \nabla^2 c\boldsymbol{B} - \dfrac{1}{c^2} \dfrac{\partial^2 c\boldsymbol{B}}{\partial t^2} = 0 \end{cases} \tag{83b}$$

We have the confirmation that $\boldsymbol{E}$ and $c\boldsymbol{B}$ fields, in regions without sources, follow a wave-like propagation at speed $c$. The consistency of equations (79) and (83b) can also be proved using Lorentz formula (43).

We now want to see what equations (83b) become when sources are present, and compare the result with equation (77). We start taking the curl of equation (56b) and the time derivative of equation (68b) and combine the results, obtaining:

$$\nabla^2 c\boldsymbol{B} - \frac{1}{c^2} \frac{\partial^2 c\boldsymbol{B}}{\partial t^2} = -\frac{1}{\varepsilon_0} \left[ \nabla \rho^{(V)} + \frac{1}{c^2} \frac{\partial \boldsymbol{j}^{(V)}}{\partial t} + \nabla \times \frac{\boldsymbol{j}^{(S)}}{c} \right] \tag{84a}$$

Similarly, if we take the curl of equation (68b) and the time derivative of equation (56b) and combine the results, we get:

$$\nabla^2 \boldsymbol{E} - \frac{1}{c^2} \frac{\partial^2 \boldsymbol{E}}{\partial t^2} = \frac{1}{\varepsilon_0} \left[ \nabla \rho^{(S)} + \frac{1}{c^2} \frac{\partial \boldsymbol{j}^{(S)}}{\partial t} - \nabla \times \frac{\boldsymbol{j}^{(V)}}{c} \right] \tag{84b}$$

It is immediate to verify that equations (84) are consistent with equation (77) thanks to the expression of Lorentz force (43), and to commutation properties of cross product and Laplacian.

*3.8. The field four-potentials*

The use of the decomposition of the force field $\boldsymbol{f}$ into the fields $\boldsymbol{E}$ and $c\boldsymbol{B}$, given by Lorentz formula, makes much easier the calculation of the force starting from field sources, as it can be seen by comparing equations (73) and (77) with equations (56) and (68). However, it is possible to further facilitate the calculation of the force adopting some mathematical tricks. Let us make reference to the field tensor $F^{\mu\nu}$. It is well known that any anti-symmetric field tensor can be represented as the sum of two anti-symmetric components: an irrotational one and a rotational one. In the case of the field tensor $F^{\mu\nu}$ we put:

$$F^{\mu\nu} = F^{(1)\mu\nu} + F^{(2)\mu\nu} \tag{85}$$

where $F^{(1)\mu\nu}$ denotes the irrotational component and $F^{(2)\mu\nu}$ the rotational one. The field tensor, being not solenoidal, can also be represented as the sum of a solenoidal component and a non-solenoidal one. The two decompositions can be made match if we succeed in imposing to the rotational component $F^{(2)\mu\nu}$ of being solenoidal.

Let us start considering the component $F^{(1)\mu\nu}$ satisfying the condition of irrotationality:

$$\varepsilon_{\kappa\lambda\mu\nu}\partial^\lambda\left(F^{(1)\mu\nu}\right)=0 \tag{86}$$

It is easy to prove that this equation admits the general integral:

$$F^{(1)\mu\nu} = \partial^\mu A^{(1)\nu} - \partial^\nu A^{(1)\mu} \tag{87}$$

where $A^{(1)\mu}$ is an arbitrary four-vector. To prove this, it is sufficient to substitute (87) into (86). The four scalar equations contained in (86) are non independent, by virtue of the identity:

$$\varepsilon_{\kappa\lambda\mu\nu}\partial^\kappa\partial^\lambda\left(F^{(1)\mu\nu}\right)=0 \tag{88}$$

Therefore the number of arbitrary scalar functions contained in the general integral (87) is equal to three. In fact it is easily recognized that its expression is left unchanged by the transformation:

$$A^{(1)\mu} = A^{(1)*\mu} + \partial^\mu \varphi^{(1)} \tag{89}$$

where $\varphi^{(1)}$ is an arbitrary scalar function. Transformation (89), that allows us to obtain the infinite solutions of our general integral from the particular solution $A^{(1)*\mu}$, is called *gauge transformation*.

Let us now consider the component $F^{(2)\mu\nu}$ whose number of independent functions must also be equal to three. Therefore we are legitimate to impose to $F^{(2)\mu\nu}$ the desired condition of solenoidality:

$$\partial_\nu\left(F^{(2)\mu\nu}\right)=0 \tag{90}$$

that implies the identity:

$$\partial_\mu\partial_\nu\left(F^{(2)\mu\nu}\right)=0 \tag{91}$$

Solenoidality of tensor $F^{(2)\mu\nu}$ assures that it can be represented as the curl of a four-vector that we indicate as $A^{(2)}{}_\alpha$:

$$F^{(2)\mu\nu} = \varepsilon^{\mu\nu\beta\alpha}\partial_\beta A^{(2)}{}_\alpha \tag{92}$$

It is easy to recognize that even this time $A^{(2)}{}_\alpha$ is defined unless the gradient of an arbitrary scalar function:

$$A^{(2)\mu} = A^{(2)*\mu} + \partial^\mu \varphi^{(2)} \tag{93}$$

Decomposition (85) therefore becomes:

$$F^{\mu\nu} = \partial^\mu A^{(1)\nu} - \partial^\nu A^{(1)\mu} + \varepsilon^{\mu\nu\beta\alpha}\partial_\beta A^{(2)}{}_\alpha \tag{94}$$

with $A^{(1)\mu}$ and $A^{(2)\mu}$ satisfying respectively equations (89) and (93). Four-vectors $A^{(1)\mu}$ and $A^{(2)\mu}$ are said *four-potentials* of the force field. We can now substitute expression (94) into the field equations (54) and (64), obtaining respectively:

$$\partial_\mu\partial^\mu A^{(1)\nu} - \partial_\mu\partial^\nu A^{(1)\mu} = \frac{1}{\varepsilon_0} J^{(S)\nu} \tag{95}$$

$$\frac{1}{2}\varepsilon^{\kappa\lambda\mu\nu}\partial_\kappa\left(\varepsilon_{\mu\nu\beta\alpha}\partial^\beta A^{(2)\alpha}\right)=\frac{1}{\varepsilon_0}J^{(V)\lambda} \tag{96}$$

It is easily recognized that:

$$\frac{1}{2}\varepsilon^{\kappa\lambda\mu\nu}\partial_\kappa\left(\varepsilon_{\mu\nu\beta\alpha}\partial^\beta A^{(2)\alpha}\right)=\partial_\mu\partial^\mu A^{(2)\lambda}-\partial_\mu\partial^\lambda A^{(2)\mu} \tag{97}$$

so that equation (96) becomes:

$$\partial_\mu\partial^\mu A^{(2)\nu}-\partial_\mu\partial^\nu A^{(2)\mu}=\frac{1}{\varepsilon_0}J^{(V)\nu} \tag{98}$$

Equations (95) and (98) are quite interesting because they show that $A^{(1)\mu}$ only depends on $J^{(S)\nu}$, while $A^{(2)\mu}$ only depends on $J^{(V)\nu}$, with great advantage for calculations. Furthermore, the gauge invariance we have found for both $A^{(1)\mu}$ and $A^{(2)\mu}$, offer us the possibility to choose, among all possible four-potentials, those depending on sources in the simplest possible way. To this end we impose the following solenoidality conditions:

$$\partial_\mu A^{(1)\mu}=0 \tag{99a}$$

$$\partial_\mu A^{(2)\mu}=0 \tag{99b}$$

referred to as *Lorentz conditions*. It is not difficult to show that potentials satisfying both conditions (99) and equations (95) and (98) always exist, and that they can be built starting from any other possible solution. In fact, considering a particular solution $A^{*\mu}$ of equation (95) or of equation (98), solenoidality condition can be written in the form:

$$\partial_\mu\left(A^{*\mu}+\partial^\mu\varphi\right)=0 \tag{100a}$$

or

$$\partial_\mu\partial^\mu\varphi=-\partial_\mu A^{*\mu} \tag{100b}$$

This is an equation in the unknown scalar $\varphi$ that is certainly soluble. We are therefore led to the following simple relations between potentials and sources:

$$\partial_\mu\partial^\mu A^{(1)\nu}=\frac{1}{\varepsilon_0}J^{(S)\nu} \tag{101a}$$

$$\partial_\mu\partial^\mu A^{(2)\nu}=\frac{1}{\varepsilon_0}J^{(V)\nu} \tag{101b}$$

In conclusion, we have found a description of the field in terms of its potentials, which share with the sources the property of being solenoidal four-vectors. While solenoidality of sources has a direct physical meaning (expressing the principle of conservation of the sources themselves), solenoidality of potentials has not a direct physical meaning, depending only on an advantageous choice in terms of computational effort, among the infinite possible ones.

In a region without sources, equations (101) become:

$$\begin{cases} \partial_\mu \partial^\mu A^{(1)\nu} = 0 \\ \partial_\mu \partial^\mu A^{(2)\nu} = 0 \end{cases} \quad (102)$$

These are the wave equations of four-potentials.

## 4. Electrodynamics

*4.1 Postulates and definitions*

We now further restrict the field of analysis, confining our attention to fields satisfying two additional postulates:

**Postulate n.3**
**Vortex sources are identically zero everywhere.**

**Postulate n. 4**
**Scalar sources enjoy the property of being charges and vice versa.**

Electrodynamics is just a force field satisfying these postulates, as well as postulates n.1 and 2. In Electrodynamics the property of matter sensible to the field and, at the same time, generator of the field, is called *electric charge*. Furthermore:

- the vector field ***E*** takes the name of *electric field*
- the axial vector field ***B*** takes the name of *magnetic field*
- the current density of scalar sources four-vector $J^{(S)\mu}$ takes the name of *electric four-current density*
- $\rho^{(S)}$ is called *electric charge density*
- $\boldsymbol{j}^{(S)}$ is called *electric current density*.

Vortex source density $\rho^{(V)}$ would have been related to magnetic monopoles, which have long been sought since the end of nineteenth century. It was pointed out by Prof. P.A.M. Dirac [9] that their existence would explain the quantization of charge in a natural way. However, up to now, no monopole has yet been seen, and probably will never be seen. In fact in this case $\rho^{(V)}$ should be related to an additional sensing property of matter, besides charge, that actually does not exist.

*4.2. The laws of Electrodynamics*

The laws of electrodynamics are given by equations (54) and (64) with $J^{(V)\mu} = 0$ so that:

$$\begin{cases} \partial_\mu F^{\mu\nu} = \dfrac{1}{\varepsilon_0} J^\nu \\ \varepsilon_{\alpha\beta\mu\nu} \partial^\beta F^{\mu\nu} = 0 \end{cases} \quad (103a)$$

or, in vector notation:

$$\begin{cases} \nabla \cdot \bm{E} = \dfrac{\rho}{\varepsilon_0} \\ \nabla \times c\bm{B} - \dfrac{1}{c}\dfrac{\partial \bm{E}}{\partial t} = \dfrac{1}{c\varepsilon_0}\bm{j} \end{cases} \qquad \begin{cases} \nabla \cdot c\bm{B} = 0 \\ \nabla \times \bm{E} + \dfrac{1}{c}\dfrac{\partial c\bm{B}}{\partial t} = 0 \end{cases} \qquad (103b)$$

where the suffix $^{(S)}$ has been dropped, failing the ambiguity with vortex sources. Equations (103) are the famous *Maxwell equations* in empty space.

Direct relationship between force and sources can be obtained from equations (73) and (77), getting:

$$\nabla \cdot \bm{f} + \frac{\bm{u}}{c} \cdot \frac{\partial \bm{f}}{\partial t} = \frac{q}{\varepsilon_0}\left[\rho - \bm{u} \cdot \frac{\bm{j}}{c}\right] \qquad (104)$$

$$\nabla^2 \bm{f} - \frac{1}{c^2}\frac{\partial^2 \bm{f}}{\partial t^2} = \frac{q}{\varepsilon_0}\left[\nabla\rho + \frac{1}{c^2}\frac{\partial \bm{j}}{\partial t}\right] - \frac{q}{\varepsilon_0}\bm{u}\times\left[\nabla\times\frac{\bm{j}}{c}\right] \qquad (105)$$

As concerns potentials, Electrodynamics only has the four-potential $A^{(1)\mu}$, being $A^{(2)\nu}$ everywhere zero by virtue of equation (101b). We can therefore drop the suffix $^{(1)}$ and rewrite equation (101a) in the form:

$$\partial_\mu \partial^\mu A^\nu = \frac{1}{\varepsilon_0} J^\nu \qquad (106)$$

Field tensor is related to the potential $A^\nu$ by the following relation, which is a particular case of equation (94):

$$F^{\mu\nu} = \partial^\mu A^\nu - \partial^\nu A^\mu \qquad (107)$$

The force acting on an electric charge $q$ moving with velocity $\bm{u}$ in an electromagnetic force field is always given by Lorentz equations (43) or (45).

All electrodynamics is contained in previous equations, on the basis of which all particular cases can be deduced.

## 5. Conclusions

Many authors have derived Maxwell's equations from electrostatics or magnetostatics and special relativity, or have derived them from analogies with Lagrangian mechanics. Here we have followed a new approach, showing that the structure of Maxwell equations comes directly from space-time structure, from the property of propagation at the speed of light and from the particular properties satisfied by the charge.

An important aspect underlined in deriving field laws was the need to define field sources. We remarked that sources can only be perceived through the field they generate, and therefore sources have to be defined in terms of field perturbations. This approach led us to adopt suitable definitions for the sources and to analyze their fundamental properties.

## References

[1] FRISCH AND WILETS: *Am. J. Phys.* **24** (1956) 574
[2] TESSMAN: *Am.J.Phys.* **34** (1966) 1048

[3] SCHWARTZ M.: *Principles of Electrodynamics.* (McGraw-Hill, New York) 1972
[4] OHANIAN H. C.: *Classical Electrodynamics.* (Allyn and Bacon, Boston) 1988
[5] KOBE D. H.: *Am J. Phys.* **54** (1986) 631-636
[6] NEUENSCHWANDER D.E., TURNER B. N.: *Am. J. Phys.* **60** (1992) 35-38
[7] OHANIAN H. C., RUFFINI R.: *Gravitation and Spacetime.* (W. W. Norton & Company) 1994
[8] MØLLER C.: *The Theory of Relativity.* (Clarendon Press, Oxford) 1972
[9] DIRAC P.A.M.: *Proc. R. Soc. London* **A144** (1931) 60; *Phys. Rev.* **74** (1948) 817